# The Comparison between the Infinitesimal Operators and Boson Operators for *SU*(3) in Cartan-Weyl Basis

Chin-Sheng Wu
Center for General Education
Yuan-Ze University, Taiwan

We present the detailed calculation of the infinitesimal operators and the boson operators for *SU* (3) in Cartan-Weyl basis. They have been used extensively as theoretical models for particle physics. We make a comparison between them, and find infinitesimal operators need modification to satisfy Cartan-Weyl basis. We also show the commutation relations and root vectors of the bases of *SL* (3,c), which displays the concise appearance and the similarity with *SU* (3).

## I. INTRODUCTION

For two decades, $q$-deformed boson operators have been applied to Hopf algebra extensively[1-3], which solves Yang-Baxter equation for many research fields such as lattice statistical mechanics, integrable quantum field theory etc. Therefore we should understand the foundation thoroughly. In this report, we present the detailed calculation of commutation relations for both the boson operators and infinitesimal operators for $SU(3)$ to see how they fit in Cartan-Weyl basis[4]. We make modification for infinitesimal operators, as Gell-Mann did[5], for Cartan-Weyl requirements. The work of $SL(3,c)$ is also displayed due to its concise appearance and the similarity with $SU(3)$.

## II. CALCULATION

Cartan-Weyl basis requires

$$[H_i, H_k] = 0, \tag{1}$$
$$[\mathbf{H}, E_\alpha] = \alpha E_\alpha, \tag{2}$$
$$[E_\alpha, E_{-\alpha}] = \alpha \bullet \mathbf{H}, \tag{3}$$
$$[E_\alpha, E_\beta] = \begin{cases} N_{\alpha\beta} E_{\alpha} +_{\beta}, & \text{if } \alpha+\beta \text{ is a root.} \\ 0, & \text{if } \alpha+\beta \text{ is not a root.} \end{cases} \tag{4}$$
$$\mathbf{H} = (H_1, \cdots, H_i), \tag{5}$$
$$\alpha = (\alpha_1, \cdots, \alpha_i), \tag{6}$$
$$\alpha \bullet \mathbf{H} = \alpha^i H_i\,. \tag{7}$$

There are eight infinitesimal operators for $SU(3)$ as follows:

$$X_1 = \begin{pmatrix} 0 & 1 & 0 \\ 1 & 0 & 0 \\ 0 & 0 & 0 \end{pmatrix}, \quad X_2 = \begin{pmatrix} 0 & -i & 0 \\ i & 0 & 0 \\ 0 & 0 & 0 \end{pmatrix}, \quad X_3 = \begin{pmatrix} 0 & 0 & 1 \\ 0 & 0 & 0 \\ 1 & 0 & 0 \end{pmatrix},$$

$$X_4 = \begin{pmatrix} 0 & 0 & -i \\ 0 & 0 & 0 \\ i & 0 & 0 \end{pmatrix}, \quad X_5 = \begin{pmatrix} 0 & 0 & 0 \\ 0 & 0 & 1 \\ 0 & 1 & 0 \end{pmatrix}, \quad X_6 = \begin{pmatrix} 0 & 0 & 0 \\ 0 & 0 & -i \\ 0 & i & 0 \end{pmatrix},$$

$$X_7 = \begin{pmatrix} 1 & 0 & 0 \\ 0 & -1 & 0 \\ 0 & 0 & 0 \end{pmatrix}, \qquad X_8 = \begin{pmatrix} \frac{1}{\sqrt{3}} & 0 & 0 \\ 0 & \frac{1}{\sqrt{3}} & 0 \\ 0 & 0 & \frac{-2}{\sqrt{3}} \end{pmatrix}.$$

The commutation relations are calculated as follows:

$$\left[ X_1 ,\ X_2 \right] = \begin{pmatrix} i & 0 & 0 \\ 0 & -i & 0 \\ 0 & 0 & 0 \end{pmatrix} - \begin{pmatrix} -i & 0 & 0 \\ 0 & i & 0 \\ 0 & 0 & 0 \end{pmatrix} = \begin{pmatrix} 2i & 0 & 0 \\ 0 & -2i & 0 \\ 0 & 0 & 0 \end{pmatrix} = 2i\, X_7 ,$$

$$\left[ X_1 ,\ X_3 \right] = \begin{pmatrix} 0 & 0 & 0 \\ 0 & 0 & 1 \\ 0 & 0 & 0 \end{pmatrix} - \begin{pmatrix} 0 & 0 & 0 \\ 0 & 0 & 0 \\ 0 & 1 & 0 \end{pmatrix} = \begin{pmatrix} 0 & 0 & 0 \\ 0 & 0 & 1 \\ 0 & -1 & 0 \end{pmatrix} = -i\, X_6 ,$$

$$\left[ X_1 ,\ X_4 \right] = \begin{pmatrix} 0 & 0 & 0 \\ 0 & 0 & -i \\ 0 & 0 & 0 \end{pmatrix} - \begin{pmatrix} 0 & 0 & 0 \\ 0 & 0 & 0 \\ 0 & i & 0 \end{pmatrix} = \begin{pmatrix} 0 & 0 & 0 \\ 0 & 0 & -i \\ 0 & -i & 0 \end{pmatrix} = -i\, X_5 ,$$

$$\left[ X_1 ,\ X_5 \right] = \begin{pmatrix} 0 & 0 & 1 \\ 0 & 0 & 0 \\ 0 & 0 & 0 \end{pmatrix} - \begin{pmatrix} 0 & 0 & 0 \\ 0 & 0 & 0 \\ 1 & 0 & 0 \end{pmatrix} = \begin{pmatrix} 0 & 0 & 1 \\ 0 & 0 & 0 \\ -1 & 0 & 0 \end{pmatrix} = -i\, X_4 ,$$

$$\left[ X_1 , X_6 \right] = \begin{pmatrix} 0 & 0 & -i \\ 0 & 0 & 0 \\ 0 & 0 & 0 \end{pmatrix} - \begin{pmatrix} 0 & 0 & 0 \\ 0 & 0 & 0 \\ i & 0 & 0 \end{pmatrix} = \begin{pmatrix} 0 & 0 & -i \\ 0 & 0 & 0 \\ -i & 0 & 0 \end{pmatrix} = -i\, X_3 ,$$

$$\left[ X_1 , X_7 \right] = \begin{pmatrix} 0 & -1 & 0 \\ 1 & 0 & 0 \\ 0 & 0 & 0 \end{pmatrix} - \begin{pmatrix} 0 & 1 & 0 \\ -1 & 0 & 0 \\ 0 & 0 & 0 \end{pmatrix} = \begin{pmatrix} 0 & -2 & 0 \\ 2 & 0 & 0 \\ 0 & 0 & 0 \end{pmatrix} = -2i\, X_2 ,$$

$$\left[ X_1 ,\ X_8 \right] = \begin{pmatrix} 0 & \frac{1}{\sqrt{3}} & 0 \\ \frac{1}{\sqrt{3}} & 0 & 0 \\ 0 & 0 & 0 \end{pmatrix} - \begin{pmatrix} 0 & \frac{1}{\sqrt{3}} & 0 \\ \frac{1}{\sqrt{3}} & 0 & 0 \\ 0 & 0 & 0 \end{pmatrix} = \begin{pmatrix} 0 & 0 & 0 \\ 0 & 0 & 0 \\ 0 & 0 & 0 \end{pmatrix},$$

$$\left[X_2, X_3\right]=\begin{pmatrix}0&0&0\\0&0&i\\0&0&0\end{pmatrix}-\begin{pmatrix}0&0&0\\0&0&0\\0&-i&0\end{pmatrix}=\begin{pmatrix}0&0&0\\0&0&i\\0&i&0\end{pmatrix}=i\,X_5,$$

$$\left[X_2, X_4\right]=\begin{pmatrix}0&0&0\\0&0&1\\0&0&0\end{pmatrix}-\begin{pmatrix}0&0&0\\0&0&0\\0&1&0\end{pmatrix}=\begin{pmatrix}0&0&0\\0&0&1\\0&-1&0\end{pmatrix}=-i\,X_6,$$

$$\left[X_2, X_5\right]=\begin{pmatrix}0&0&-i\\0&0&0\\0&0&0\end{pmatrix}-\begin{pmatrix}0&0&0\\0&0&0\\i&0&0\end{pmatrix}=\begin{pmatrix}0&0&-i\\0&0&0\\-i&0&0\end{pmatrix}=-i\,X_3,$$

$$\left[X_2, X_6\right]=\begin{pmatrix}0&0&-1\\0&0&0\\0&0&0\end{pmatrix}-\begin{pmatrix}0&0&0\\0&0&0\\-1&0&0\end{pmatrix}=\begin{pmatrix}0&0&-1\\0&0&0\\1&0&0\end{pmatrix}=i\,X_4,$$

$$\left[X_2, X_7\right]=\begin{pmatrix}0&i&0\\i&0&0\\0&0&0\end{pmatrix}-\begin{pmatrix}0&-i&0\\-i&0&0\\0&0&0\end{pmatrix}=\begin{pmatrix}0&2i&0\\2i&0&0\\0&0&0\end{pmatrix}=2i\,X_1,$$

$$\left[X_2, X_8\right]=\begin{pmatrix}0&-\frac{i}{\sqrt{3}}&0\\\frac{i}{\sqrt{3}}&0&0\\0&0&0\end{pmatrix}-\begin{pmatrix}0&\frac{-i}{\sqrt{3}}&0\\\frac{i}{\sqrt{3}}&0&0\\0&0&0\end{pmatrix}=\begin{pmatrix}0&0&0\\0&0&0\\0&0&0\end{pmatrix},$$

$$\left[X_3, X_4\right]=\begin{pmatrix}i&0&0\\0&0&0\\0&0&-i\end{pmatrix}-\begin{pmatrix}-i&0&0\\0&0&0\\0&0&i\end{pmatrix}=\begin{pmatrix}2i&0&0\\0&0&0\\0&0&-2i\end{pmatrix}=i\,X_7+\sqrt{3}\;i\,X_8,$$

$$\left[X_3, X_5\right]=\begin{pmatrix}0&1&0\\0&0&0\\0&0&0\end{pmatrix}-\begin{pmatrix}0&0&0\\1&0&0\\0&0&0\end{pmatrix}=\begin{pmatrix}0&1&0\\-1&0&0\\0&0&0\end{pmatrix}=-\imath\,X_2,$$

$$\left[X_3, X_6\right]=\begin{pmatrix}0&\imath&0\\0&0&0\\0&0&0\end{pmatrix}-\begin{pmatrix}0&0&0\\-\imath&0&0\\0&0&0\end{pmatrix}=\begin{pmatrix}0&\imath&0\\\imath&0&0\\0&0&0\end{pmatrix}=\imath\,X_1,$$

$$[X_3, X_7] = \begin{pmatrix} 0 & 0 & 0 \\ 0 & 0 & 0 \\ 1 & 0 & 0 \end{pmatrix} - \begin{pmatrix} 0 & 0 & 1 \\ 0 & 0 & 0 \\ 0 & 0 & 0 \end{pmatrix} = \begin{pmatrix} 0 & 0 & -1 \\ 0 & 0 & 0 \\ 1 & 0 & 0 \end{pmatrix} = \imath X_4,$$

$$[X_3, X_8] = \begin{pmatrix} 0 & 0 & -\frac{2}{\sqrt{3}} \\ 0 & 0 & 0 \\ \frac{2}{\sqrt{3}} & 0 & 0 \end{pmatrix} - \begin{pmatrix} 0 & 0 & \frac{1}{\sqrt{3}} \\ 0 & 0 & 0 \\ -\frac{2}{\sqrt{3}} & 0 & 0 \end{pmatrix} = \begin{pmatrix} 0 & 0 & -\sqrt{3} \\ 0 & 0 & 0 \\ \sqrt{3} & 0 & 0 \end{pmatrix} = -\sqrt{3}\, \imath X_4,$$

$$[X_4, X_5] = \begin{pmatrix} 0 & -\imath & 0 \\ 0 & 0 & 0 \\ 0 & 0 & 0 \end{pmatrix} - \begin{pmatrix} 0 & 0 & 0 \\ \imath & 0 & 0 \\ 0 & 0 & 0 \end{pmatrix} = \begin{pmatrix} 0 & -\imath & 0 \\ -\imath & 0 & 0 \\ 0 & 0 & 0 \end{pmatrix} = -\imath X_1,$$

$$[X_4, X_6] = \begin{pmatrix} 0 & 1 & 0 \\ 0 & 0 & 0 \\ 0 & 0 & 0 \end{pmatrix} - \begin{pmatrix} 0 & 0 & 0 \\ 1 & 0 & 0 \\ 0 & 0 & 0 \end{pmatrix} = \begin{pmatrix} 0 & 1 & 0 \\ -1 & 0 & 0 \\ 0 & 0 & 0 \end{pmatrix} = -\imath X_2,$$

$$[X_4, X_7] = \begin{pmatrix} 0 & 0 & 0 \\ 0 & 0 & 0 \\ \imath & 0 & 0 \end{pmatrix} - \begin{pmatrix} 0 & 0 & -\imath \\ 0 & 0 & 0 \\ 0 & 0 & 0 \end{pmatrix} = \begin{pmatrix} 0 & 0 & \imath \\ 0 & 0 & 0 \\ 0 & 0 & 0 \end{pmatrix} = \imath X_3,$$

$$[X_4, X_8] = \begin{pmatrix} 0 & 0 & \frac{2\imath}{\sqrt{3}} \\ 0 & 0 & 0 \\ \frac{\imath}{\sqrt{3}} & 0 & 0 \end{pmatrix} - \begin{pmatrix} 0 & 0 & -\frac{2}{\sqrt{3}} \\ 0 & 0 & 0 \\ -\frac{2\imath}{\sqrt{3}} & 0 & 0 \end{pmatrix} = \begin{pmatrix} 0 & 0 & \sqrt{3}\,\imath \\ 0 & 0 & 0 \\ \sqrt{3}\,\imath & 0 & 0 \end{pmatrix} = \sqrt{3}\, \imath X_3,$$

$$[X_5, X_6] = \begin{pmatrix} 0 & 0 & 0 \\ 0 & \imath & 0 \\ 0 & 0 & -\imath \end{pmatrix} - \begin{pmatrix} 0 & 0 & 0 \\ 0 & -\imath & 0 \\ 0 & 0 & \imath \end{pmatrix} = \begin{pmatrix} 0 & 0 & 0 \\ 0 & 2\imath & 0 \\ 0 & 0 & -2\imath \end{pmatrix} = -i X_7 + \sqrt{3}\, \imath X_8,$$

$$[X_5, X_7] = \begin{pmatrix} 0 & 0 & 0 \\ 0 & 0 & 0 \\ 0 & -1 & 0 \end{pmatrix} - \begin{pmatrix} 0 & 0 & 0 \\ 0 & 0 & -1 \\ 0 & 0 & 0 \end{pmatrix} = \begin{pmatrix} 0 & 0 & 0 \\ 0 & 0 & 1 \\ 0 & -1 & 0 \end{pmatrix} = -\imath X_6,$$

$$[X_5, X_8] = \begin{pmatrix} 0 & 0 & 0 \\ 0 & 0 & -\frac{2}{\sqrt{3}} \\ 0 & \frac{1}{\sqrt{3}} & 0 \end{pmatrix} - \begin{pmatrix} 0 & 0 & 0 \\ 0 & 0 & \frac{1}{\sqrt{3}} \\ 0 & -\frac{2}{\sqrt{3}} & 0 \end{pmatrix} = \begin{pmatrix} 0 & 0 & 0 \\ 0 & 0 & -\sqrt{3} \\ 0 & \sqrt{3} & 0 \end{pmatrix} = -\sqrt{3}\, i X_6,$$

$$[X_6, X_7] = \begin{pmatrix} 0 & 0 & 0 \\ 0 & 0 & 0 \\ 0 & -i & 0 \end{pmatrix} - \begin{pmatrix} 0 & 0 & 0 \\ 0 & 0 & i \\ 0 & 0 & 0 \end{pmatrix} = \begin{pmatrix} 0 & 0 & 0 \\ 0 & 0 & -i \\ 0 & -i & 0 \end{pmatrix} = -i X_5,$$

$$[X_6, X_8] = \begin{pmatrix} 0 & 0 & 0 \\ 0 & 0 & \frac{2}{\sqrt{3}}i \\ 0 & \frac{i}{\sqrt{3}} & 0 \end{pmatrix} - \begin{pmatrix} 0 & 0 & 0 \\ 0 & 0 & -\frac{i}{\sqrt{3}} \\ 0 & -\frac{2i}{\sqrt{3}} & 0 \end{pmatrix} = \begin{pmatrix} 0 & 0 & 0 \\ 0 & 0 & \sqrt{3}i \\ 0 & \sqrt{3}i & 0 \end{pmatrix} = \sqrt{3}\, i X_5,$$

$$[X_7, X_8] = \begin{pmatrix} \frac{1}{\sqrt{3}} & 0 & 0 \\ 0 & -\frac{1}{\sqrt{3}} & 0 \\ 0 & 0 & 0 \end{pmatrix} - \begin{pmatrix} \frac{1}{\sqrt{3}} & 0 & 0 \\ 0 & -\frac{1}{\sqrt{3}} & 0 \\ 0 & 0 & 0 \end{pmatrix} = \begin{pmatrix} 0 & 0 & 0 \\ 0 & 0 & 0 \\ 0 & 0 & 0 \end{pmatrix}.$$

Eight boson operators, which corresponding to infinitesimal operators, are given by

$$H_1 = \frac{1}{2\sqrt{3}}(n_1 - n_2) \qquad \Leftrightarrow X_7,$$

$$H_2 = \frac{1}{6}(n_1 + n_2 - 2n_3) \qquad \Leftrightarrow X_8,$$

$$E_\alpha = \frac{1}{\sqrt{6}}(C_1^+ C_3) \qquad \Leftrightarrow X_3,$$

$$E_{-\alpha} = \frac{1}{\sqrt{6}} C_3^+ C_1 \qquad \Leftrightarrow X_4,$$

$$E_\beta = \frac{1}{\sqrt{6}} C_3^+ C_2 \qquad \Leftrightarrow X_5,$$

$$E_{-\beta} = \frac{1}{\sqrt{6}} C_2^+ C_3 \qquad \Leftrightarrow X_6,$$

$$E_\gamma = \frac{1}{\sqrt{6}} C_1^+ C_2 \qquad \Leftrightarrow X_1,$$

$$E_{-\gamma} = \frac{1}{\sqrt{6}} C_2^+ C_1 \qquad \Leftrightarrow X_2.$$

The commutation relations are calculated as follows:

$$\left[ E_\gamma , E_{-\gamma} \right] = \left[ \frac{1}{\sqrt{6}} C_1^+ C_2 , \frac{1}{\sqrt{6}} C_2^+ C_1 \right]$$

$$= \frac{1}{6} \left( C_1^+ C_2 C_2^+ C_1 - C_2^+ C_1 C_1^+ C_2 \right)$$

$$= \frac{1}{6} \left\{ C_1^+ C_1 \left( 1 + C_2^+ C_2 \right) - C_1 C_1^+ C_2^+ C_2 \right\}$$

$$= \frac{1}{6} \left\{ \left( C_1^+ C_1 - C_1 C_1^+ \right) C_2^+ C_2 + C_1^+ C_2 \right\}$$

$$= \frac{1}{6} \left( n_1 - n_2 \right)$$

$$= \frac{1}{\sqrt{3}} H_1 ,$$

$$\left[ E_\gamma , E_\alpha \right] = \left[ \frac{1}{\sqrt{6}} C_1^+ C2, \frac{1}{\sqrt{6}} C_1^+ C_3 \right]$$

$$= \frac{1}{6} \left( C_1^+ C_2 C_1^+ C_3 - C_1^+ C_3 C_1^+ C_2 \right)$$

$$= 0,$$

$$\left[ E_\gamma , E_{-\alpha} \right] = \left[ \frac{1}{6} C_1^+ C_2 , \frac{1}{\sqrt{6}} C_3^+ C_1 \right] = \frac{1}{6} ( C_1^+ C_2 C_3^+ C_1 - C_3^+ C_1 C_1^+ C_2 )$$

$$= \frac{1}{6} C_2 C_3^+ ( C_1^+ C_1 - C_1 C_1^+ ) = - \frac{1}{\sqrt{6}} E_\beta ,$$

$$\left[ E_\gamma , E_\beta \right] = \left[ \frac{1}{\sqrt{6}} C_1^+ C_2 , \frac{1}{\sqrt{6}} C_3^+ C_2 \right]$$

$$= \frac{1}{6} \left( C_1^+ C_2 C_3^+ C_2 - C_3^+ C_2 C_1^+ C_2 \right)$$

$$= \frac{1}{6} \left( C2 \, C_3^+ - C_3^+ C2 \right) C_1^+ C2$$

$$= 0,$$

$$\left[ E_\gamma , E_{-\beta} \right] = \left[ \frac{1}{\sqrt{6}} C_1^+ C_2 , \frac{1}{6} C_2^+ C_3 \right]$$

$$= \frac{1}{6} \left( C_1^+ C_2 C_2^+ C_3 - C_2^+ C_3 C_1^+ C_2 \right)$$

$$= \frac{1}{6} \left\{ C_1^+ C_3 \left( 1 + C_2^+ C_2 \right) - C_2^+ C_3 C_1^+ C_2 \right\}$$

$$= \frac{1}{6} \left( C_1^+ C_3 + C_1^+ C_3 C_2^+ C_2 - C_2^+ C_3 C_1^+ C_2 \right)$$

$$= \frac{1}{\sqrt{6}} E_\alpha ,$$

$$
\begin{aligned}
\left[E_\gamma , H_1\right] &= \left[\frac{1}{6} C_1^+ C_2 , \frac{1}{2\sqrt{3}} (n_1 - n_2)\right] \\
&= \left(\frac{1}{6} C_1^+ C_2 \frac{1}{2\sqrt{3}} \left(n_1 - n_2\right) - \frac{1}{2\sqrt{3}} \frac{1}{6} \left(n_1 - n_2\right) C_1^+ C_2\right) \\
&= \frac{1}{\sqrt{2}6} \left\{ C_1^+ C_2 n_1 - C_1^+ C_2 n_2 - n_1 C_1^+ C_2 + n_2 C_1^+ C_2 \right\} \\
&= \frac{1}{\sqrt{2}6} \left\{ \left( C_1^+ n_1 - n_1 C_1^+ \right) C_2 - \left( C_1^+ C_2 n_2 - C_1^+ n_2 C_2 \right) \right\} \\
&= \frac{1}{\sqrt{2}6} \left\{ - C_1^+ C_2 + C_1^+ \left( - C_2 n_2 + n_2 C_2 \right) \right\} \\
&= \frac{1}{\sqrt{2}} \frac{1}{6} \left\{ - C_1^+ C_2 + C_1^+ \left( - C_2 \right) \right\} \\
&= -\frac{1}{\sqrt{3}} E_\gamma ,
\end{aligned}
$$

$$
\begin{aligned}
\left[E_\gamma , H_2\right] &= \left[ \frac{1}{\sqrt{6}} C_1^+ C_2 , \frac{1}{6} \left(n_1 + n_2 - 2n_3 \right) \right] \\
&= \frac{1}{\sqrt{6}6} \left\{ C_1^+ C_2 \left(n_1 + n_2 - 2n_3\right) - \left(n_1 + n_2 - 2n_3 \right) C_1^+ C_2 \right\} \\
&= \frac{1}{\sqrt{6}6} \left\{ C_1^+ C_2 n_1 - n_1 C_1^+ C_2 + C_1^+ C_2 n_2 - n_2 C_1^+ C_2 - 2C_1^+ C_2 n_3 + 2n_3 C_1^+ C_2 \right\} \\
&= \frac{1}{\sqrt{6}} \frac{1}{6} \left\{ - C_1^+ C_2 + C_1^+ C_2 \right\} \\
&= 0,
\end{aligned}
$$

$$
\begin{aligned}
\left[E_{-\gamma} , E_\alpha\right] &= \left[ \frac{1}{\sqrt{6}} C_2^+ C_1 , \frac{1}{\sqrt{6}} C_1^+ C_3 \right] \\
&= \frac{1}{6} \left( C_2^+ C_1 C_1^+ C_3 - C_1^+ C_3 C_2^+ C_1 \right) \\
&= \frac{1}{6} \left\{ C_2^+ C_3 \left( 1 + C_1^+ C_1 \right) - C_3 C_2^+ C_1^+ C_1 \right\} \\
&= \frac{1}{6} \left\{ C_2^+ C_3 + C_2^+ C_3 C_1^+ C_1 - C_3 C_2^+ C_1^+ C_1 \right\} \\
&= \frac{1}{6} C_2^+ C_3 = \frac{1}{\sqrt{6}} E_{-\beta} ,
\end{aligned}
$$

$$
\begin{aligned}
\left[E_{-\gamma}, E_{-\alpha}\right] &= \left[\frac{1}{\sqrt{6}} C_2^+ C_1 , \frac{1}{\sqrt{6}} C_3^+ C_1\right] \\
&= \frac{1}{6}\left(C_2^+ C_1 C_3^+ C_1 - C_3^+ C_1 C_2^+ C_1\right) = 0, \\
\left[E_{-\gamma}, E_{\beta}\right] &= \left[\frac{1}{6} C_2^+ C_1 , C_3^+ C_2\right] \\
&= \frac{1}{6}\left(C_2^+ C_1 C_3^+ C_2 - C_3^+ C_2 C_2^+ C_1\right) \\
&= \frac{1}{6}\left\{\left(C_2 C_2^+ - 1\right) C_1 C_3^+ - C_2 C_2^+ C_3^+ C_1\right\} \\
&= \frac{1}{6}\left\{- C_1 C_3^+ + C_2 C_2^+ \left(C_1 C_3^+ - C_3^+ C_1\right)\right\} \\
&= -\frac{1}{\sqrt{6}} E_{-\alpha},
\end{aligned}
$$

$$
\begin{aligned}
\left[E_{-\gamma}, E_{-\beta}\right] &= \left[\frac{1}{\sqrt{6}} C_2^+ C_1 , \frac{1}{\sqrt{6}} C_2^+ C_3\right] \\
&= \frac{1}{6}\left(C_2^+ C_1 C_2^+ C_3 - C_2^+ C_3 - C_2^+ C_1\right) \\
&= 0,
\end{aligned}
$$

$$
\begin{aligned}
\left[E_{-\gamma}, H_1\right] &= \left[\frac{1}{\sqrt{6}} C_2^+ C_1 , \frac{1}{2\sqrt{3}}\left(n_1 - n_2\right)\right] \\
&= \frac{1}{6\sqrt{2}}\left\{C_2^+ C_1 \left(n_1 - n_2\right) - \left(n_1 - n_2\right) C_2^+ C_1\right\} \\
&= \frac{1}{6\sqrt{2}}\left\{C_2^+ C_1 n_1 - C_2^+ C_1 n_2 - n_1 C_2^+ C_1 + n_2 C_2^+ C_1\right\} \\
&= \frac{1}{6\sqrt{2}}\left\{C_2^+ \left(C_1 n_1 - n_1 C_1\right) - C_1 \left(C_2^+ n_2 - n_2 C_2^+\right)\right\} \\
&= \frac{1}{6\sqrt{2}}\left\{C_2^+ C_1 + C_1 C_2^+\right\} \\
&= \frac{2}{\sqrt{6}} \frac{2}{\sqrt{2}} E_{-\gamma} \\
&= \frac{1}{\sqrt{3}} E_{-\gamma},
\end{aligned}
$$

$$
\begin{aligned}
\left[E_{-\gamma}\,, H_2\right] &= \left[\frac{1}{\sqrt{6}}C_2^+\,C_1\,, \frac{1}{6}\left(n_1+n_2-2n_3\right)\right] \\
&= \frac{1}{6\sqrt{6}}\left\{C_2^+\,C_1\left(n_1+n_2-2n_3\right)-\left(n_1+n_2-2n_3\right)C_2^+\,C_1\right\} \\
&= \frac{1}{6\sqrt{6}}\left\{C_2^+\,C_1\,n_1-n_1\,C_2^+\,C_1+C_2^+\,C_1\,n_2-n_2\,C_2^+\,C_1-2C_2^+\,C_1\,n_3+2n_3\,C_2^+\,C_1\right\} \\
&= \frac{1}{6\sqrt{6}}\left\{C_2^+\,C_1-C_2^+\,C_1\right\} \\
&= 0,
\end{aligned}
$$

$$
\begin{aligned}
\left[E_{\alpha}\,, E_{-\alpha}\right] &= \left[\frac{1}{\sqrt{6}}C_1^+\,C_3\,, \frac{1}{\sqrt{6}}C_3^+\,C_1\right] \\
&= \frac{1}{6}\left\{C_1^+\,C_3\,C_3^+\,C_1-C_3^+\,C_1\,C_1^+\,C_3\right\} \\
&= \frac{1}{6}\left\{\left(-1+C_1\,C_1^+\right)\left(1+C_3^+\,C_3\right)-C_1\,C_1^+\,C_3^+\,C_3-C_1\,C_1^+\,C_3^+\,C_3\right\} \\
&= \frac{1}{6}\left\{-1-C_3^+\,C_3+C_1\,C_1^++C_1\,C_1^+\,C_3^+\,C_3-C_1\,C_1^+\,C_3^+\,C_3\right\} \\
&= \frac{1}{6}\left\{-1-n_3+1+n_1\right\} \\
&= \frac{1}{6}\left(n_1-n_3\right)=\frac{1}{2\sqrt{3}}H_1+\frac{1}{2}H_2,
\end{aligned}
$$

$$
\begin{aligned}
\left[E_{\alpha}, E_{\beta}\right] &= \left[\frac{1}{\sqrt{6}}C_1^+\,C_3\,, \frac{1}{\sqrt{6}}C_3^+\,C_2\right] = \frac{1}{6}\left(C_1^+\,C_3\,C_3^+\,C_2-C_3^+\,C_2\,C_1^+\,C_3\right) \\
&= \frac{1}{6}C_1^+\left(1+C_3^+\,C_3\right)C_2-C_3^+\,C_2\,C_1^+\,C_2 \\
&= \frac{1}{6}\left(C_1^+\,C_2+C_1^+\,C_2\,C_3^+\,C_2-C_1^+\,C_2\,C_3^+\,C_3\right) \\
&= \frac{1}{\sqrt{6}}E_{\gamma},
\end{aligned}
$$

$$
\begin{aligned}
\left[E_{\alpha}, E_{-\beta}\right] &= \left[\frac{1}{\sqrt{6}}C_1^+\,C_3\,, \frac{1}{\sqrt{6}}C_2^+\,C_3\right] \\
&= \frac{1}{6}\left(C_1^+\,C_3\,C_2^+\,C_3-C_2^+\,C_3\,C_1^+\,C_3\right) = 0,
\end{aligned}
$$

$$[E_\alpha, H_1] = \left[ \frac{1}{\sqrt{6}} C_1^+ C_3 , \frac{1}{2\sqrt{3}} ( n_1 - n_2 ) \right]$$

$$= \frac{1}{6\sqrt{2}} \left\{ C_1^+ C_3 ( n_1 - n_2 ) - (n_1 - n_2) C_1^+ C_3 \right\}$$

$$= \frac{1}{6\sqrt{2}} \left\{ C_1^+ C_3 n_1 - n_1 C_1^+ C_3 - C_1^+ C_3 n_2 + n_2 C_1^+ C_3 \right\}$$

$$= \frac{1}{6\sqrt{2}} \left\{ -C_1^+ C_3 \right\} = \frac{-1}{2\sqrt{3}} E_\alpha,$$

$$[E_\alpha , H_2] = \left[ \frac{1}{\sqrt{6}} C_1^+ C_3 , \frac{1}{6} ( n_1 + n_2 - 2n_3 ) \right]$$

$$= \frac{1}{6\sqrt{6}} \left\{ C_1^+ C_3 ( n_1 + n_2 - 2n_3 - n_1 + n_2 - 2n ) \right\}$$

$$= \frac{1}{6\sqrt{6}} \left\{ C_1^+ C_3 n_1 - n_1 C_1^+ C_3 + C_1^+ C_3 n_2 - n_2 C_1^+ C_3 - 2C_1^+ C_3 n_3 + 2n_3 C_1^+ C_3 \right\}$$

$$= \frac{1}{6\sqrt{6}} \left\{ -C_1^+ C_3 + 0 - 2C_1^+ C_3 \right\} = -\frac{1}{2} E_\alpha,$$

$$[E_{-\alpha} , E_\beta] = \left[ \frac{1}{\sqrt{6}} C_3^+ C_1 , \frac{1}{\sqrt{6}} C_3^+ C_2 \right] = \frac{1}{6} \left[ C_3^+ C_1 C_3^+ C_2 - C_3^+ C_2 C_3^+ C_1 \right] = 0,$$

$$[E_{-\alpha} , E_{-\beta}] = \left[ \frac{1}{\sqrt{6}} C_3^+ C_1 , \frac{1}{\sqrt{6}} C_2^+ C_3 \right] = \frac{1}{6} \left( C_3^+ C_1 C_2^+ C_3 - C_2^+ C_3 C_3^+ C_1 \right)$$

$$= \frac{1}{6} \left\{ C_3^+ C_3 \left( C_2^+ C_1 \right) - C_2^+ C_1 \left( 1 + C_3^+ C_3 \right) \right\}$$

$$= \frac{1}{6} \left\{ -C_2^+ C_1 \right\}$$

$$= -\frac{1}{\sqrt{6}} E_{-\gamma},$$

$$[E_{-\alpha} , H_1 ] = \left[ \frac{1}{\sqrt{6}} C_3^+ C_1 , \frac{1}{2\sqrt{3}} ( n_1 - n_2 ) \right]$$

$$= \frac{1}{6\sqrt{2}} \left( C_3^+ C_1 ( n_1 - n_2 ) - ( n_1 - n_2 ) C_3^+ C_1 \right)$$

$$= \frac{1}{6\sqrt{2}} \left\{ C_3^+ C_1 n_1 - n_1 C_3^+ C_1 - C_3^+ C_1 n_2 + n_2 C_3^+ C_1 \right\}$$

$$= \frac{1}{6\sqrt{2}} \left\{ C_3^+ C_1 \right\} = \frac{1}{2\sqrt{3}} E_{-\alpha},$$

$$
\begin{aligned}
[E_{-\alpha}\,, H_2] &= \left[\ \frac{1}{\sqrt{6}} C_3^+\ C_1\,, \frac{1}{6}(\ n_1 + n_2 - 2n_3\ )\right] \\
&= \frac{1}{6\sqrt{6}}\left\{ C_3^+\ C_1\ (\ n_1 + n_2 - 2n_3\ ) - (\ n_1 + n_2 - 2n_3\ ) C_3^+\ C_1\ \right\} \\
&= \frac{1}{6\sqrt{6}}\left\{ C_3^+\ C_1\ n_1 - n_1\ C_3^+\ C_1 - 2C_3^+\ C_1\ n_3 + 2n_3\ C_3^+\ C_1\ \right\} \\
&= \frac{1}{6\sqrt{6}}\left\{ C_3^+\ C_1 + 2C_1\ C_3^+\ \right\} \ = \ \frac{1}{6}\{E_{-\alpha} + 2E_{-\alpha}\ \} \ = \ \frac{1}{2}E_{-\alpha},
\end{aligned}
$$

$$
\begin{aligned}
[E_{\beta}\,, E_{-\beta}] &= \left[\ \frac{1}{6} C_3^+\ C_2\,, \frac{1}{\sqrt{6}} C_2^+\ C_3\ \right] \ = \ \frac{1}{6}\left\{ C_3^+\ C_2\ C_2^+\ C_3 - C_2^+\ C_3\ C_3^+\ C_2 \right\} \\
&= \frac{1}{6}\left\{ (\ C_3\ C_3^+ - 1)(1 + C_2^+\ C_2\ ) - C_2^+\ C_2\ C_3\ C_3^+\ \right\} \\
&= \frac{1}{6}\left\{ C_3\ C_3^+ - C_2^+\ C_2 - 1 + C_3\ C_3^+\ C_2^+\ C_2 - C_2^+\ C_2\ C_3\ C_3^+\ \right\} \\
&= \frac{1}{6}\left\{ C_3^+\ C_3 - 1 - C_2^+\ C_2 - 1 \right\} \ = \ \frac{1}{6}\{\ n_3 - n_2\ \} = \frac{1}{2\sqrt{3}} H_1 - \frac{1}{2} H_2,
\end{aligned}
$$

$$
\begin{aligned}
[E_{\beta}\,, H_1] &= \left[\ \frac{1}{\sqrt{6}} C_3^+\ C_2\,, \frac{1}{2\sqrt{3}}(\ n_1 - n_2\ )\right] \\
&= \frac{1}{6\sqrt{2}}\left( C_3^+\ C_2\ (\ n_1 - n_2) - (\ n_1 - n_2\ ) C_3^+\ C_2 \right) \\
&= \frac{1}{6\sqrt{2}}\left\{ C_3^+\ C_2\ n_1 - n_1\ C_3^+\ C_2 - C_3^+\ C_2\ n_2 + n_2\ C_3^+\ C_2 \right\} \\
&= \frac{1}{6\sqrt{2}}\left\{ C_3^+\ (- C_2 n_2 + n_2 C_2\ ) \right\} \ = \ \frac{-1}{6\sqrt{2}} C_3^+\ C_2 \ = \ -\frac{1}{2\sqrt{3}} E_{\beta},
\end{aligned}
$$

$$
\begin{aligned}
[E_{\beta}\,, H_2\ ] &= \left[\ \frac{1}{\sqrt{6}} C_3^+\ C_2\,, \frac{1}{6}(\ n_1 + n_2 - 2n_3\ )\right] \\
&= \frac{1}{6\sqrt{\ }}\left\{ C_3^+\ C_2\ (\ n_1 + n_2 - 2n_3\ ) - (\ n_1 + n_2 - 2n_3\ ) C_3^+\ C_2\ \right\} \\
&= \frac{1}{6\sqrt{6}}\left\{ C_3^+\ C_2\ n_2 - n_2\ C_3^+\ C_2 - 2C_3^+\ C_2\ n_3 + 2n_3\ C_3^+\ C_2\ \right\} \\
&= \frac{1}{6\sqrt{6}}\left\{ C_3^+\ C_2 + 2C_2\ C_3^+\ \right\} \ = \ \frac{1}{2} E_{\beta},
\end{aligned}
$$

$$
\begin{aligned}
\left[E_{-\beta}\,, H_1\,\right] &= \left[\,\frac{1}{\sqrt{6}}C_2^+\,C_3\,,\frac{1}{2\sqrt{3}}(\,n_1-n_2\,)\right] \\
&= \frac{1}{6\sqrt{2}}\left(C_2^+\,C_3\,(\,n_1-n_2)-(n_1-n_2)C_2^+\,C_3\,\right) \\
&= \frac{1}{6\sqrt{2}}\left\{\left(C_2^+\,C_3\,n_1-n_1\,C_2^+\,C_3\,\right)-C_2^+\,C_3\,n_2+n_2\,C_2^+\,C_3\,\right\} \\
&= \frac{1}{6\sqrt{2}}\left\{0-C_2^+\,n_2\,C_3+n_2\,C_2^+\,C_3\,\right\} \\
&= \frac{1}{6\sqrt{2}}C_2^+\,C_3 \\
&= \frac{1}{2\sqrt{3}}E_{-\beta}\,, \\
\left[E_{-\beta}\,, H_2\,\right] &= \left[\,\frac{1}{\sqrt{6}}C_2^+\,C_3\,,\frac{1}{6}(\,n_1+n_2-2n_3\,)\right] \\
&= \frac{1}{6\sqrt{6}}\left\{C_2^+\,C_3\,(\,n_1+n_2-2n_3\,)-(\,n_1+n_2-2n\,)C_2^+\,C_3\,\right\} \\
&= \frac{1}{6\sqrt{6}}\left\{C_2^+\,C_3\,n_2-n_2\,C_2^+\,C_3-2C_2^+\,C_3\,n_3+2n_3\,C_2^+\,C_3\,\right\} \\
&= \frac{1}{6\sqrt{6}}\left\{-C_2^+\,C_3+2C_2^+\,(-C_3)\right\} \\
&= \frac{-3}{6\sqrt{6}}C_2^+\,C_3 \\
&= -\frac{1}{2}E_{-\beta}\,, \\
\left[H_1\,, H_2\,\right] &= \left[\,\frac{1}{2\sqrt{3}}(\,n_1-n_2),\frac{1}{6}(\,n_1+n_2-2n_3\,)\right] \\
&= 0.
\end{aligned}
$$

## III. RESULTS

The commutation relations of infinitesimal operators and corresponding boson operators are listed as below.

$$[\,X_1\,, X_2\,] = 2\,i\,X_7 \qquad\qquad [\,E_\gamma\,, E_{-\gamma}\,] = \frac{1}{\sqrt{3}}H_1$$

$$[\,X_1\,, X_3\,] = -i\,X_6 \qquad\qquad [\,E_\gamma\,,\; E_\alpha\,] = 0$$

$$[\,X_1\,, X_4\,] = -i\,X_5 \qquad\qquad [E_\gamma, E_{-\alpha}] = -\frac{1}{\sqrt{6}}E_\beta$$

$[X_1, X_5] = -i X_4$ $\qquad$ $[E_\gamma, E_\beta] = 0$

$[X_1, X_6] = -i X_3$ $\qquad$ $[E_\gamma, E_{-\beta}] = \frac{1}{\sqrt{6}} E_\alpha$

$[X_1, X_7] = -2 i X_2$ $\qquad$ $[E_\gamma, H_1] = -\frac{1}{\sqrt{3}} E_\gamma$

$[X_1, X_8] = 0$ $\qquad$ $[E_\gamma, H_2] = 0$

$[X_2, X_3] = i X_5$ $\qquad$ $[E_{-\gamma}, E_\alpha] = \frac{1}{\sqrt{6}} E_{-\beta}$

$[X_2, X_4] = -i X_6$ $\qquad$ $[E_{-\gamma}, E_{-\alpha}] = 0$

$[X_2, X_5] = -i X_3$ $\qquad$ $[E_{-\gamma}, E_\beta] = -\frac{1}{\sqrt{6}} E_{-\alpha}$

$[X_2, X_6] = i X_4$ $\qquad$ $[E_{-\gamma}, E_{-\beta}] = 0$

$[X_2, X_7] = 2i X_1$ $\qquad$ $[E_{-\gamma}, H_1] = \frac{1}{\sqrt{3}} E_{-\gamma}$

$[X_2, X_8] = 0$ $\qquad$ $[E_{-\gamma}, H_2] = 0$

$[X_3, X_4] = i X_7 + \sqrt{3}\, i X_8$ $\qquad$ $[E_\alpha, E_{-\alpha}] = \frac{1}{2\sqrt{3}} H_1 + \frac{1}{2} H_2$

$[X_3, X_5] = -i X_2$ $\qquad$ $[E_\alpha, E_\beta] = \frac{1}{\sqrt{6}} E_\gamma$

$[X_3, X_6] = -i X_1$ $\qquad$ $[E_\alpha, E_{-\beta}] = 0$

$[X_3, X_7] = i X_4$ $\qquad$ $[E_\alpha, H_1] = -\frac{1}{2\sqrt{3}} E_\alpha$

$[X_3, X_8] = -\sqrt{3}_i\, X_4$ $\qquad$ $[E_\alpha, H_2] = -\frac{1}{2} E_\alpha$

$[X_4, X_5] = -i X_1$ $\qquad$ $[E_{-\alpha}, E_\beta] = 0$

$[X_4, X_6] = -i X_2$ $\qquad$ $[E_{-\alpha}, E_{-\beta}] = \frac{1}{\sqrt{6}} E_{-\gamma}$

$[X_4, X_7] = i X_3$ $\qquad$ $[E_{-\alpha}, H_1] = \frac{1}{2\sqrt{3}} E_{-\alpha}$

$[X_4, X_8] = \sqrt{3}\, i X_3$ $\qquad$ $[E_{-\alpha}, H_2] = \frac{1}{2} E_{-\alpha}$

$[X_5, X_6] = -i X_7 + \sqrt{3}\, i X_8$ $\qquad$ $[E_\beta, E_{-\beta}] = \frac{1}{2\sqrt{3}} H_1 - \frac{1}{2} H_2$

$[X_5, X_7] = -i X_6$ $\qquad$ $[E_\beta, H_1] = -\frac{1}{2\sqrt{3}} E_\beta$

$$[X_5, X_8] = -\sqrt{3}\, i\, X_6 \qquad [E_\beta, H_2] = \frac{1}{2} E_\beta$$

$$[X_6, X_7] = -i\, X_5 \qquad [E_{-\beta}, H_1] = \frac{1}{2\sqrt{3}} E_{-\beta}$$

$$[X_6, X_8] = \sqrt{3}\, i\, X_5 \qquad [E_{-\beta}, H_2] = -\frac{1}{2} E_{-\beta}$$

$$[X_7, X_8] = 0 \qquad [H_1, H_2] = 0$$

Root vectors *α, β, γ* for the infinitesimal operators and boson operators

| | | | |
|---|---|---|---|
| ***X₁*** | (-2,0) | $\gamma$ | $(-\frac{1}{\sqrt{3}},0)$ |
| ***X₂*** | (2,0) | $-\gamma$ | $(\frac{1}{\sqrt{3}},0)$ |
| ***X₃*** | $(1,-\sqrt{3})$ | $\alpha$ | $(-\frac{1}{2\sqrt{3}},-\frac{1}{2})$ |
| ***X₄*** | $(1,\sqrt{3})$ | $-\alpha$ | $(\frac{1}{2\sqrt{3}},\frac{1}{2})$ |
| ***X₅*** | $(-1,-\sqrt{3})$ | $\beta$ | $(-\frac{1}{2\sqrt{3}},\frac{1}{2})$ |
| ***X₆*** | $(-1,\sqrt{3})$ | $-\beta$ | $(\frac{1}{2\sqrt{3}},-\frac{1}{2})$ |

## IV. COMPARISON AND MODIFICATION

From above calculated results, we found the boson operators meet all the requirements of Cartan-Weyl basis. However the infinitesimal operators do not satisfy the equation (4). For instance, $[X_1, X_5] = -iX_4$, which is supposed to be zero according to the addition of root vectors. The contradiction is probably due to the nature of matrix multiplication, which results nonzero product for matrix with these forms. Gell-Mann introduced the combination of bases such as $X_1 \pm iX_2, X_3 \pm iX_4, X_5 \pm iX_6$, which are no longer elements of $SU(3)$. By choosing these combination bases, one can see that they are exactly the infinitesimal operators[6] of $SL(3,c)$. All the requirements of Cartan-Weyl basis are satisfied. Because $SL(3,c)$ displays the instructive information about the root vectors, we present here for comparison with different and normalized $X_7$ later.

$$H_1 = \begin{pmatrix} 1 & 0 & 0 \\ 0 & -1 & 0 \\ 0 & 0 & 0 \end{pmatrix} \qquad H_2 = \begin{pmatrix} 0 & 0 & 0 \\ 0 & 1 & 0 \\ 0 & 0 & -1 \end{pmatrix}$$

$$E_\alpha = \begin{pmatrix} 0 & 1 & 0 \\ 0 & 0 & 0 \\ 0 & 0 & 0 \end{pmatrix} \qquad E_{-\alpha} = \begin{pmatrix} 0 & 0 & 0 \\ 1 & 0 & 0 \\ 0 & 0 & 0 \end{pmatrix}$$

$$E_\beta = \begin{pmatrix} 0 & 0 & 0 \\ 0 & 0 & 1 \\ 0 & 0 & 0 \end{pmatrix} \qquad E_{-\beta} = \begin{pmatrix} 0 & 0 & 0 \\ 0 & 0 & 0 \\ 0 & 1 & 0 \end{pmatrix}$$

$$E_\gamma = \begin{pmatrix} 0 & 0 & 1 \\ 0 & 0 & 0 \\ 0 & 0 & 0 \end{pmatrix} \qquad E_{-\gamma} = \begin{pmatrix} 0 & 0 & 0 \\ 0 & 0 & 0 \\ 1 & 0 & 0 \end{pmatrix}$$

$$[H_1, H_2] = 0$$

$$[H_1, H_\alpha] = 2E_\alpha \qquad [H_1, E_{-\alpha}] = -2E_{-\alpha}$$

$$[H_2, H_\alpha] = -E_\alpha \qquad [H_2, E_{-\alpha}] = E_{-\alpha}$$

$$[H_1, E_\beta] = -E_\beta \qquad [H_1, E_{-\beta}] = E_{-\beta}$$

$$\lfloor H_2, E_\beta \rfloor = 2E_\beta \qquad \lfloor H_2, E_{-\beta} \rfloor = -2E_{-\beta}$$

$$[H_1, E_\gamma] = E_\gamma \qquad [H_1, E_{-\gamma}] = -E_{-\gamma}$$

$$[H_2, E_\gamma] = E_\gamma \qquad [H_2, E_{-\gamma}] = -E_{-\gamma}$$

$$\lfloor E_\alpha, E_{-\alpha} \rfloor = H_1$$

$$[E_\beta, E_{-\beta}] = H_2$$

$$\lfloor E_\gamma,\ E_{-\gamma} \rfloor = H_1 + H_2$$

$$[E_\alpha , E_\beta] = E_\gamma \qquad [E_{-\alpha} , E_{-\beta}] = -E_\gamma$$

$$[E_\alpha , E_{-\beta}] = 0 \qquad [E_\beta , E_{-\alpha}] = 0$$

$$[E_\alpha , E_\gamma] = 0 \qquad [E_{-\alpha} , E_{-\gamma}] = 0$$

$$[E_\beta , E_\gamma] = 0 \qquad [E_{-\beta} , E_{-\gamma}] = 0$$

$$[E_\beta , E_{-\gamma}] = E_{-\alpha} \qquad [E_\gamma , E_{-\beta}] = E_\alpha$$

$$[E_\alpha , E_{-\gamma}] = -E_{-\beta} \qquad [E_\gamma , E_{-\alpha}] = -E_\beta$$

$$E_\alpha \ (2,-1) \qquad E_{-\alpha}(-2,1)$$

$$E_\beta \ (-1,2) \qquad E_{-\beta} \ (1,-2)$$

$$E_\gamma \ (1,1) \qquad E_{-\gamma} \ (-1,-1)$$

One notices that there are two different vector lengths and if we draw the diagram of these vectors, the symmetric axis of the diagram is $x = y$. We now choose the different basis for $X_7$, as Gell-Mann did, and normalize it.

$$X_7 = \frac{1}{\sqrt{3}} \begin{pmatrix} 1 & 0 & 0 \\ 0 & 1 & 0 \\ 0 & 0 & -2 \end{pmatrix}$$

We obtain the Gell-Mann's root vectors

$\boldsymbol{X_1}$ (2,0)
$\boldsymbol{X_2}$ (-2,0)
$\boldsymbol{X_3}$ (-1,$\sqrt{3}$)
$\boldsymbol{X_4}$ (1,-$\sqrt{3}$)
$\boldsymbol{X_5}$ (1,$\sqrt{3}$)
$\boldsymbol{X_6}$ (-1,-$\sqrt{3}$)

The equation (4) is satisfied, and it is used successfully in quark model. It's marvelous to learn that the two different sources, infinitesimal operators and boson operators, demonstrate the same Cartan-Weyl basis.